\documentclass[10pt,journal,compsoc]{IEEEtran}

\ifCLASSOPTIONcompsoc
  \usepackage[nocompress]{cite}
\else
  \usepackage{cite}
\fi
\usepackage{amsmath}
\usepackage{stfloats}
\usepackage[colorlinks=true, citecolor=blue, linkcolor=blue]{hyperref}   
\usepackage{graphicx}
\usepackage{xcolor}

\begin{document}
%
\title{Holographic Microscopy with Python and HoloPy}

%
%

\author{Solomon Barkley,
  Thomas G. Dimiduk,
  Jerome Fung,
  David M. Kaz,
  Vinothan N. Manoharan,
  Ryan McGorty,
  Rebecca W. Perry,
  and Anna Wang
  \IEEEcompsocitemizethanks{
    \IEEEcompsocthanksitem Harvard John A. Paulson School of Engineering
    and Applied Sciences and Department of Physics, Harvard University,
    Cambridge, MA 02138.\protect\\
  E-mail: vnm@seas.harvard.edu
  \IEEEcompsocthanksitem Thomas G. Dimiduk is currently at Tesla, Inc.
  \IEEEcompsocthanksitem Jerome Fung is currently at the Department of
  Physics, Wellesley
  College.
  \IEEEcompsocthanksitem David M. Kaz is currently at Agilent
  Technologies.
  \IEEEcompsocthanksitem Ryan McGorty is currently at the Department of
  Physics and Biophysics, University of San Diego.
    \IEEEcompsocthanksitem Rebecca W. Perry is currently at Wayfair, Inc.
  \IEEEcompsocthanksitem Anna Wang is currently at the Department of
  Molecular Biology, Massachusetts General Hospital.}
  }

%
%

\IEEEtitleabstractindextext{%
  \begin{abstract}
    A holographic microscope captures interference patterns, or
    holograms, that encode three-dimensional (3D) information about the
    object being viewed. Computation is essential to extracting that 3D
    information. By wrapping low-level scattering codes and taking
    advantage of Python's data analysis ecosystem, HoloPy makes it easy
    for experimentalists to use modern, sophisticated inference methods
    to analyze holograms. The resulting data can be used to understand
    how small particles or microorganisms move and interact. The project
    illustrates how computational tools can enable experimental methods
    and new experiments.
  \end{abstract}
  
  }

\maketitle

%
\IEEEpeerreviewmaketitle

\IEEEraisesectionheading{\section{Introduction}\label{sec:introduction}}

%
%
%
%

\begin{figure}
	\centering
	\includegraphics{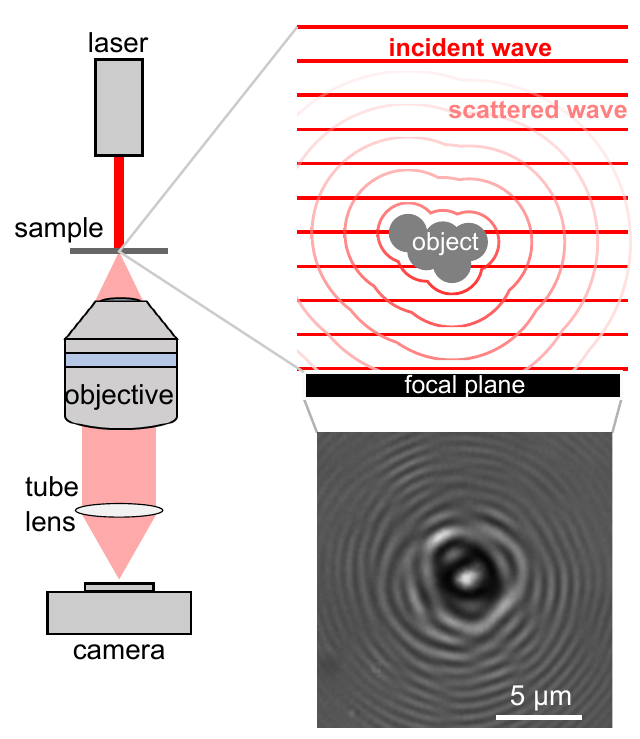}
	\caption{An in-line holographic microscope (left) consists of a laser,
    an objective lens, and a camera. Some of the incident light is
    scattered by the sample and interferes with the transmitted light
    (upper right). The interference pattern recorded at the focal plane
    is called a hologram (lower right).}
	\label{fig:holographic_microscope}
\end{figure}

\IEEEPARstart{H}{olographic} microscopy is an elegant way to image tiny
objects in three dimensions. In its simplest form, a holographic
microscope is the same as a conventional light microscope, except that
the light source is a laser. The coherent light from the laser scatters
from the sample and interferes with the transmitted light, producing a
pattern of bright and dark fringes called a \emph{hologram}, as shown in
Figure~\ref{fig:holographic_microscope} (although colloquially the term
``hologram'' often refers to a 3D image, we use it in its original sense
of a 2D interference pattern). Traditionally, holograms were recorded on
film, and the 3D information encoded in their fringes was
\emph{reconstructed} by shining light through the hologram. Gabor, the
discoverer of holography, showed that reconstruction yields a 3D image
because it reproduces the original field scattered from the sample,
including both its amplitude and phase~\cite{gabor_new_1948}.

\begin{figure*}
	\centering
	\includegraphics{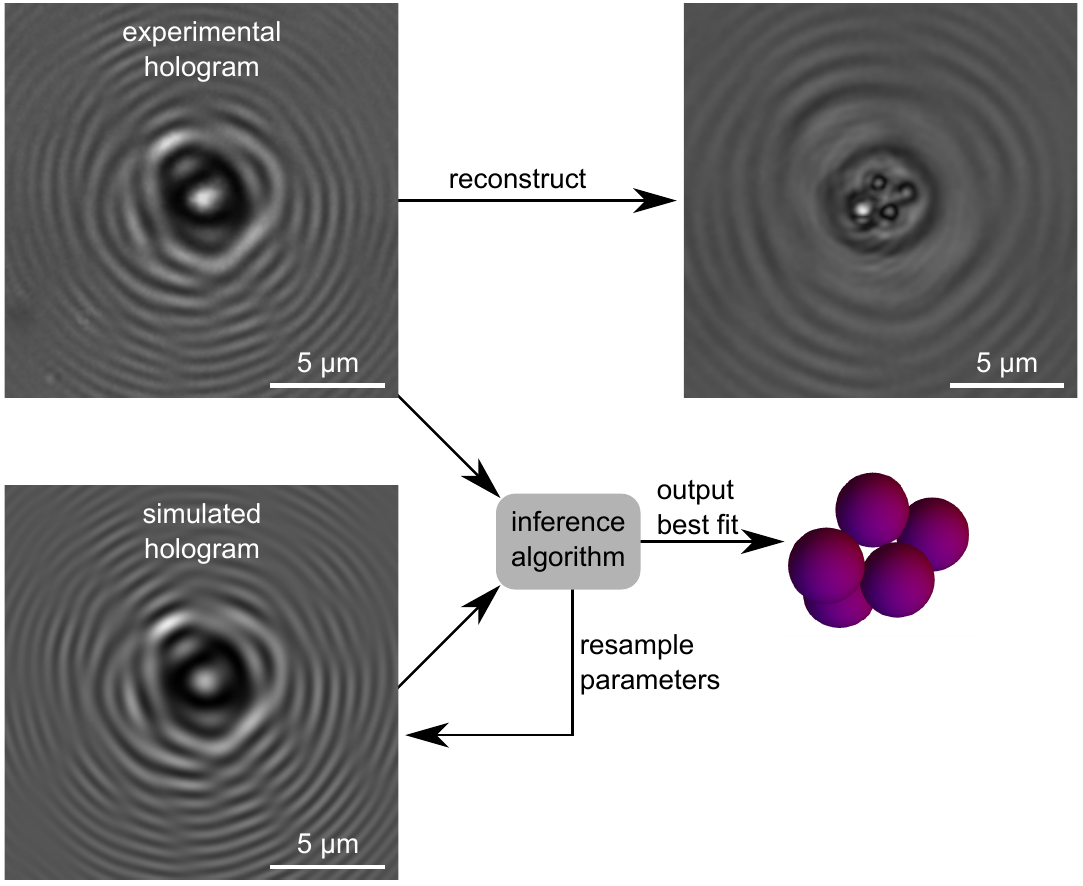}
	\caption{Two computational methods to extract 3D information from an
    in-line hologram. Reconstruction (top) involves simulating light
    propagating through the hologram to yield a 3D image of the sample.
    From the reconstruction, parameters of interest, such as the
    positions of individual particles, can be measured by processing the
    image. This method tends to introduce artifacts such as blurring and
    fringe noise, which can be seen in the reconstructed image.
    Inference (bottom) involves iteratively simulating holograms and
    comparing them to data to extract information about the object
    imaged by the microscope.}
	\label{fig:holo_process}
\end{figure*}

Nowadays holograms are recorded digitally, and the 3D information is
reconstructed by simulating light shining through the image
(Figure~\ref{fig:holo_process}). Digital recording and reconstruction
have enabled many scientific advances. For example, holographic
time-series (or movies) recorded by high-speed digital cameras have been
reconstructed to reveal how micro-organisms swim in three
dimensions~\cite{lewis2006swimming} and how tracer particles move in a
turbulent flow~\cite{sheng2008using}. Our research group does similar
experiments. We record holograms of tiny colloidal particles or
biological organisms moving in a fluid to understand how they organize
themselves into more complex structures. To do that, we must determine
their 3D positions in each hologram of a time-series.

This task is complicated by two problems inherent to reconstruction.
First, the reconstructed image can have unphysical
artifacts~\cite{gabor_microscopy_1949, pu2003intrinsic, pu2004intrinsic,
  cheong_strategies_2010}, such as the blurring and fringes shown in
Figure~\ref{fig:holo_process}. Second, the reconstruction must be
further processed to extract information about the object, adding
another layer of analysis.

A computational method has recently emerged that circumvents both of
these problems. Starting with a model that can \emph{exactly simulate}
(to within numerical precision) the light scattered from microscopic
objects, an algorithm adjusts the parameters in the model until the
simulation matches, or fits, a measured
hologram~\cite{ovryn_imaging_2000, lee_characterizing_2007}. Using this
technique, we can infer the 3D positions of multiple objects directly
from the hologram, without ever reconstructing the scattered field
(Figure~\ref{fig:holo_process}). Though computationally more demanding,
this \emph{inference} approach avoids the artifacts inherent to
reconstruction while still taking advantage of the phase information
encoded by the holograms.

The inference approach has transformed how our group does experiments.
Previously, we needed to design the microscope so that the artifacts
were minimized in 3D reconstructions. Now that we no longer rely on
reconstruction, we can keep the instrument simple---no more than a
conventional light microscope equipped with a laser and a digital
camera---and use computers to handle the complex task of extracting the
information we want.

Twelve years ago, we set out to make that complex task as simple as
possible. We developed HoloPy, a Python package that simulates
scattering and fits scattering models to measured holograms. Here we
describe how Python made it possible to create an
experimentalist-friendly package, and how we use it to analyze data from
experiments in soft-matter physics and biology.

\section{Why Python?}

We originally wrote HoloPy to numerically reconstruct holograms from our
experiments. At the time (2006), many image processing codes were
written in IDL or MATLAB. We chose Python because it was (and is) a
general-purpose, cross-platform, free and open-source language; because
we could write a numerical reconstruction routine in just a few lines of
code using the NumPy package\cite{oliphant06:_numpy}; and because it was
easy for us to learn. The last criterion was especially important for a
burgeoning research group, since we needed to spend most of our time
building instruments, making measurements, and thinking about
physics---and not, in general, writing code.

Since we started using inference instead of reconstruction to analyze
holograms, the package has changed a lot. But the focus on making it
easy to use and modify has not. HoloPy is still written with
experimental scientists in mind. Python has served us well in pursuit of
this goal, because it allows us to wrap legacy codes in a common
interface and to integrate existing Python packages to add
functionality.

\subsection{Wrapping scattering codes}
\label{wrap_scat_codes}

When we moved to an inference framework, we needed HoloPy to simulate
scattering from different kinds of objects. There were two
possibilities: implement scattering models directly in Python or use
existing implementations written in Fortran. It wasn't easy to write
scattering models from scratch. The scattering from our samples is
complicated because the size of the particles is comparable to the
wavelength of light (about half a micrometer). Even the simplest
scattering model---the Lorenz-Mie theory, which describes how a
homogeneous sphere scatters light---requires calculating a host of
special functions and evaluating series expansions with dozens of terms,
all the while avoiding overflow or underflow
errors\cite{wiscombe_improved_1980}. We realized that there was no point
in writing new code in Python when there were already well-tested
Fortran codes that did all of these tasks efficiently. And there were
codes for many different kinds of objects, including not only spheres
but also collections of spheres, spheroids, cylinders, and even
arbitrarily shaped particles.

However, each scattering code had its own coordinate system and
interface, which were often cumbersome: parameters were specified by
input files in arcane formats or through command-line arguments. So we
used Python to wrap the codes in an object-oriented interface with a
common interface. The f2py system made it easy to wrap Fortran
subroutines. For codes designed to run as executables rather than as
subroutines, we wrote setup scripts to compile the executables and
wrappers to communicate with them through Python's subprocess module,
which works across all platforms. With this approach, new users needed
to learn only one interface, which we could make intuitive and concise
(see Figure~\ref{fig:scat_calc}).

\subsection{Integration with data analysis packages}

As we added more complex scattering models to HoloPy, we found that we
needed to include more sophisticated methods to fit those models to the
data. Fitting complex models involves varying many free parameters. A
scattering model for a cluster of five colloidal particles, like that
shown in Figure~\ref{fig:holo_process}, might have 15 free parameters,
corresponding to the positions of all the particles. Finding the
parameters that best fit a hologram of this system involves a search in
a 15-dimensional space.

To address this problem, we integrated a Python implementation for
Markov-chain Monte Carlo (MCMC) calculations,
emcee~\cite{foreman2013emcee}, with HoloPy. MCMC is a powerful method
for fitting models with many parameters to data, and emcee is an
implementation that can handle many different kinds of problems.
Integrating emcee with HoloPy allowed us to define a concise interface
and to provide sensible defaults specifically for analysis of holograms.
It was easy to integrate the two packages, since both were written in
Python.

Although packages like emcee didn't drive us to start working in Python,
we now benefit from Python's growing prominence in the data science
community. There are now many packages for MCMC and visualization that
we can use to enhance HoloPy.

\section{A brief tour of HoloPy}

Here we show some examples of calculations in HoloPy to illustrate how
the scattering models and inference frameworks are integrated.

\subsection{Simulating a hologram from a given object}

\begin{figure*}
	\centering
	\includegraphics{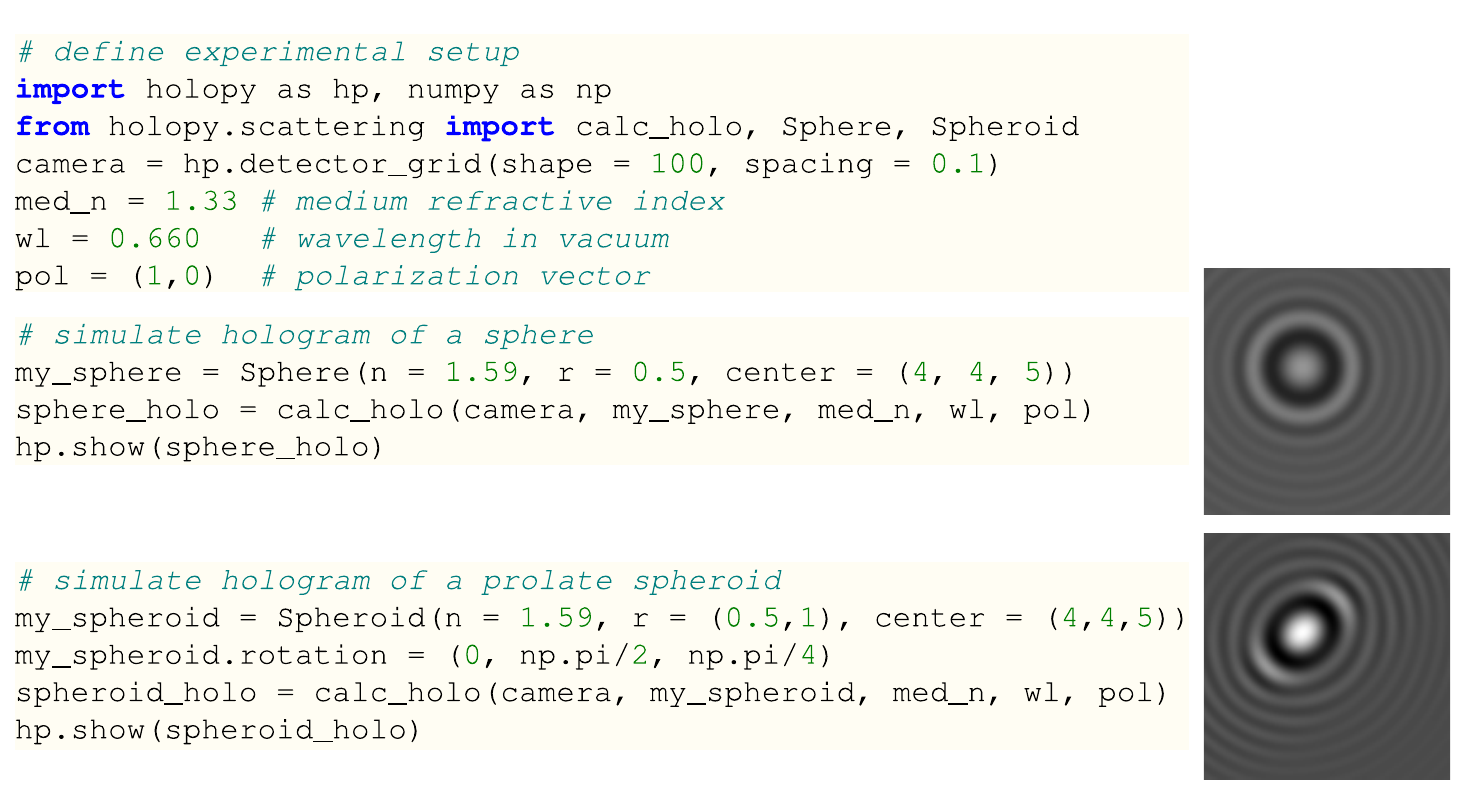}
	\caption{HoloPy code to simulate holograms produced by a 1-$\mu$m
    polystyrene sphere and a 2$\times$1-$\mu$m ellipsoid, both in water.
    Simulated holograms are shown at right. Note that HoloPy is agnostic
    to the units of length, as long as they are consistent. Here, all
    lengths are in $\mu$m.}
	\label{fig:scat_calc}
\end{figure*}

A scattering model is essential to the inference approach.
Figure~\ref{fig:scat_calc} shows how to simulate a hologram using
HoloPy's built-in models. We first tell HoloPy about the shape of our
camera, the size of its pixels, and the properties of the incident
illumination---its wavelength, polarization, and the refractive index of
the medium it travels in. Then we define an object. In the example, we
define a spherical particle using five parameters: its center position
($x,y,z$), radius $r$, and refractive index $n$. Given this information,
HoloPy calculates the scattered field at the camera. It then numerically
calculates the interference between the scattered and transmitted waves
to simulate the hologram.

Simulations like these are a convenient way to explore how holograms
change for different particles or different parameters. For example,
it's easy to change the particle from a sphere to an ellipsoid, as shown
in Figure~\ref{fig:scat_calc}. We don't have to specify the scattering
model when we do this. We define our object, and HoloPy automatically
selects the appropriate model: the Lorenz-Mie theory for the sphere, and
the T-Matrix formulation~\cite{mishchenko_electromagnetic_2003} for the
ellipsoid.

\subsection{Inferring an object's properties from a hologram}

\begin{figure*}
	\centering
	\includegraphics{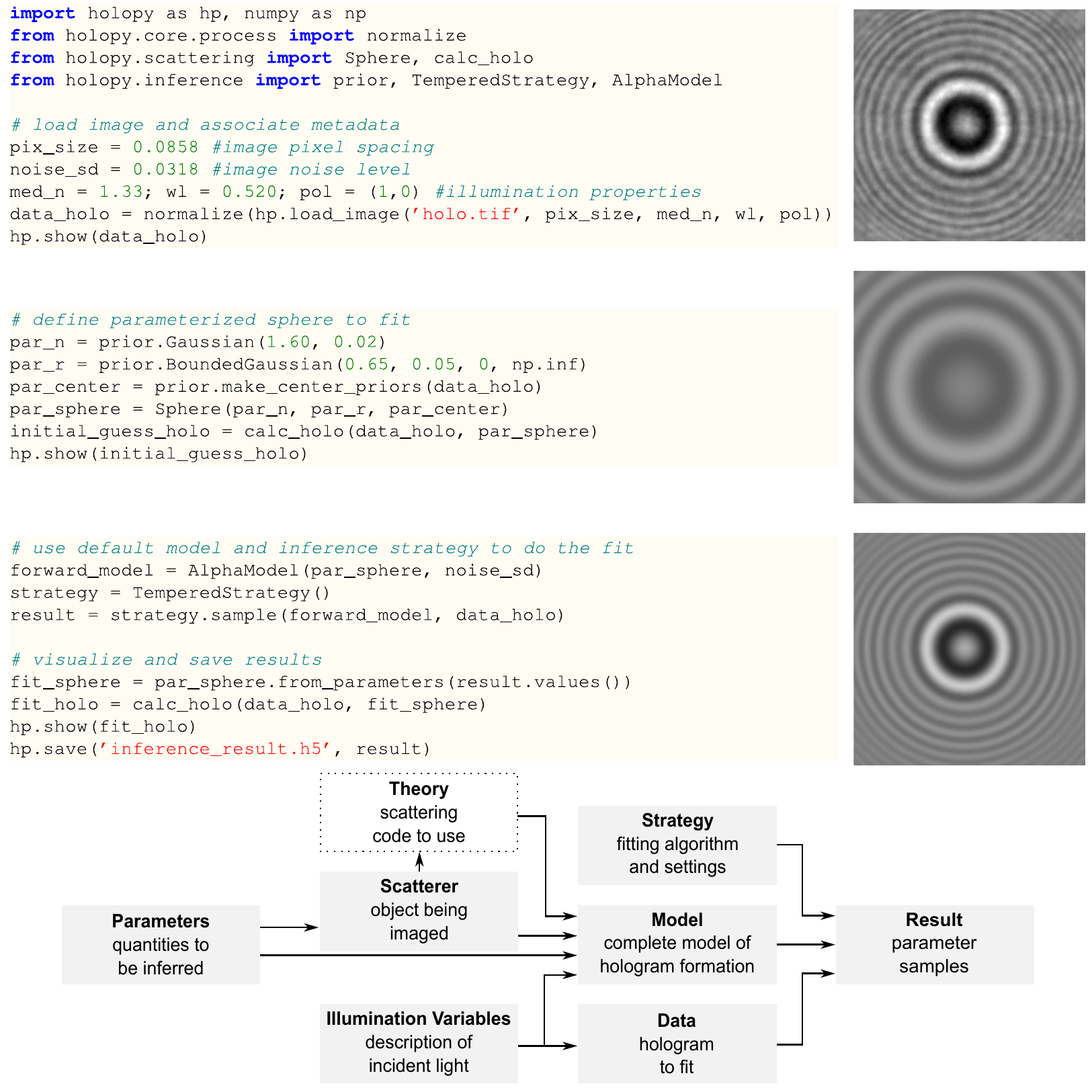}
	\caption{Example code and diagram for a HoloPy inference calculation.
    The measured hologram is produced by a 1.3-$\mu$m polystyrene
    particle in water. Starting from this hologram, an initial guess,
    and a model for hologram formation, HoloPy determines the parameters
    that best fit the data. The data, initial guess, and best-fit
    holograms are shown at right. The relationships between objects
    (bottom) makes the inference calculation modular.}
	\label{fig:inference_calc}
\end{figure*}

The real power of simulation lies in solving the inference (or inverse)
problem: given a measured hologram, determine the position and
properties of a microscopic particle. A sample inference calculation is
shown in Figure~\ref{fig:inference_calc}. We specify the following: (1)
variables describing the incident illumination; (2) an experimentally
measured hologram (\texttt{data\_holo}) and an estimate of the noise in
that hologram; (3) the expected shape of the particle (here, a
\texttt{Sphere}), along with parameters of the particle to vary, and
over what ranges we expect them to vary; (4) an initial guess for the
free parameters; (5) a complete model for the simulation, which includes
a scattering theory (which we let HoloPy choose automatically) and a
model for the optical train of the microscope (here called
\texttt{AlphaModel}~\cite{lee_characterizing_2007}); (6) an inference
algorithm, along with settings for that algorithm (here, we use
emcee and let HoloPy choose the default settings from
\texttt{TemperedStrategy()}). With this information, HoloPy determines the
best fit, as shown in Figure~\ref{fig:inference_calc}. It also outputs a
results object that contains a full description of the model and the
inference strategy as well as the fitted parameter values.

Although what happens behind the scenes is complex, the code in
Figure~\ref{fig:inference_calc} is not. In keeping with our goal of
making complex tasks simple, HoloPy chooses intelligent defaults for the
MCMC algorithm and the scattering model. Also, the object hierarchy
makes the calculation modular. It is straightforward to change, for
example, the inference strategy while keeping the scattering model
fixed, or vice versa.

\section{Experiment meets computation}

We've given only a taste of HoloPy above. Additional functionality to
manage and analyze experimentally acquired holograms is central to the
package. Here we explain how we designed this functionality to meet the
needs of experimentalists.

\subsection{Enabling sophisticated data analysis}

In a typical experiment we quantify not only the positions of particles,
but also the uncertainty on those positions. The uncertainty is
primarily related to the noise in the hologram. We care about the
uncertainty because we ultimately want to understand how a particle
moves or interacts, and the uncertainty tells us how much to trust our
measurements.

\begin{figure}
	\centering
	\includegraphics{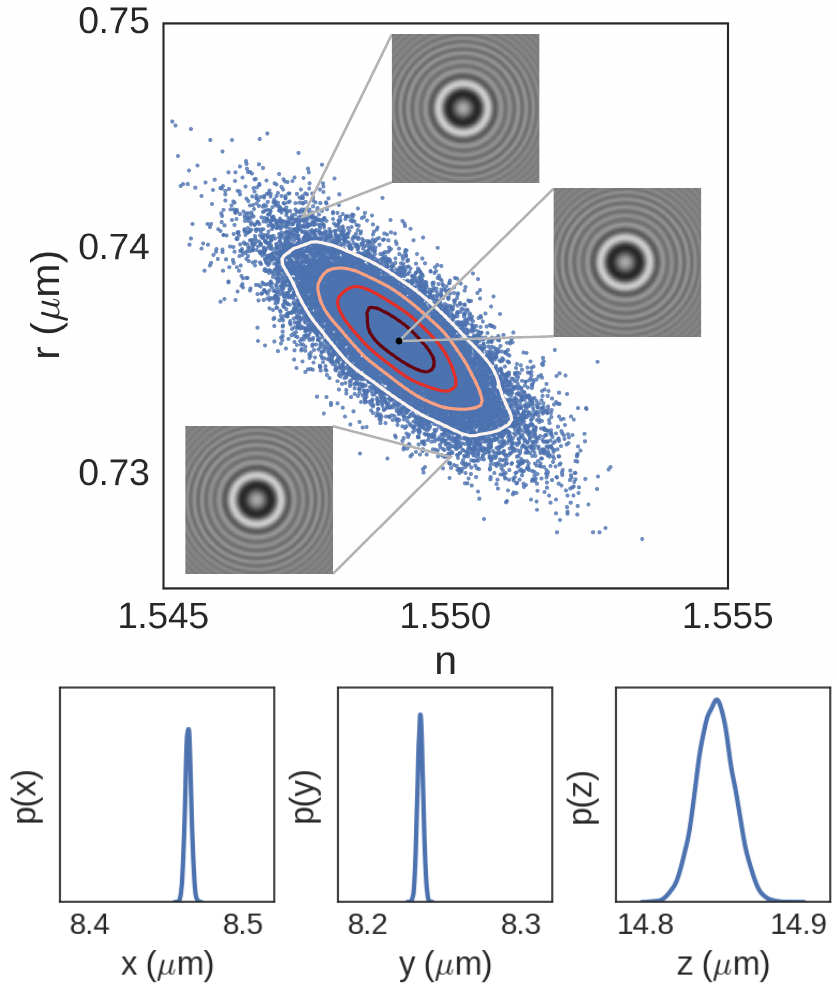}
	\caption{An MCMC calculation returns many sets of parameter values,
    each corresponding to a simulated hologram. Here we show
    visualizations of the MCMC samples returned from analysis of the
    hologram shown in Figure~\ref{fig:inference_calc}. The joint
    distribution at top provides a measure of the correlations between
    fitted parameters, while the fully marginalized distributions at
    bottom characterize the uncertainties of each parameter in isolation
    (bottom). The top diagram also shows the contour lines of a kernel
    density estimate of the joint distribution, as well as the simulated
    holograms corresponding to the best fit (center) and two other
    samples.}
  	\label{fig:distributions}
\end{figure}

HoloPy uses a Bayesian framework to determine the uncertainty. Given the
scattering model, measured hologram, and estimate of the noise in that
hologram, it constructs a posterior probability distribution for the
free parameters in the model~\cite{dimiduk_bayesian_2016}. Because the
posterior is a function of many parameters, it's inefficient to
calculate it on a regular grid spanning the whole parameter space.
Instead, the MCMC framework randomly samples sets of parameters from the
posterior. For each set of parameters (or ``sample''), HoloPy simulates
the hologram and calculates the difference between the simulated and
measured holograms. The MCMC algorithm uses this difference to choose
the next sample. The distribution of all the samples tells us the
best-fit parameters as well as their uncertainties and covariances
(Figure~\ref{fig:distributions}).

The Bayesian framework is particularly well-suited to experiments on
microscopic particles, for which properties such as the size or
refractive index are usually known only approximately. The most
convenient way to specify this ``fuzzy'' knowledge is through a
probability distribution, or prior. In the example in
Figure~\ref{fig:inference_calc}, we use a Gaussian prior for the
refractive index of the sphere and a bounded Gaussian prior for the
radius.

Because the MCMC analysis returns all the samples, and not just the
best-fit parameters, we can visualize and characterize the posterior
probability density. In Figure~\ref{fig:distributions}, we plot the
samples for the refractive index ($n$) and radius ($r$) from an MCMC
analysis of a hologram of a single sphere. Each point represents a
particular set of parameters---or, equivalently, a simulated hologram.
From these samples, we can construct distributions using histograms or
kernel density estimates with the seaborn package
(\url{https://seaborn.pydata.org}). The plots allow us to characterize
the uncertainty. We see, for example, that our estimates of $n$ are
correlated with our estimates of $r$, which tells us that a measured
hologram is equally well fit by a small radius and large index as by a
large radius and small index. That makes sense physically: the effective
size of the particle that the light ``sees'' is related to the product
of $n$ and $r$. We can also construct marginal distributions for single
parameters such as $x$, $y$, or $z$. These plots give us the uncertainty
in each parameter, after marginalizing (accounting for) the
uncertainties and correlations among all the other parameters. With this
wealth of information about the uncertainties, we can optimize the
experiment or propagate the uncertainties to test physical models of how
the particles move.

MCMC methods are powerful tools for Bayesian inference, but they come
at a cost. A typical MCMC calculation might involve thousands of
samples, which means HoloPy has to simulate thousands of holograms for
each measured one. The number of samples is the price we must pay for
accurate uncertainty estimates.

That price can be made smaller through a few computational tricks that
we built into HoloPy. Simulating a 512$\times$512-pixel hologram
involves making 262,144 individual scattering calculations, which takes
(in total) about a second on a modern CPU. Since the MCMC algorithm does
thousands of simulations for each fit, we realized that we could save a
significant amount of computational time by reducing the number of
pixels in each MCMC step. The simplest way to do this is to choose a
random set of pixels in the measured hologram and fit just those pixels,
rather than the whole hologram\cite{dimiduk_random-subset_2014}. We
found that we could get accurate results with only tens of pixels, which
speeds up the calculation by a factor of 100 or more. This strategy
works because the hologram contains a lot of redundant information in
its fringes. We tell HoloPy to use random subsetting through the
\texttt{Strategy} object.

These are just a few ways in which HoloPy enables the kinds of data
analysis that experimentalists need. While some of the functionality,
such as random subsetting, is built into HoloPy, the rest comes from the
broader Python ecosystem, including packages like emcee and seaborn.

\subsection{Time-series analysis}

Python's extended infrastructure also comes in handy for time-series
analysis. We record time-series because we are interested not only in
the size and shape of microscopic particles, but also in understanding
how they move. Tracking the motion of a bacterium tells us about how it
searches for food~\cite{wang_tracking_2016}, and tracking the motion of
colloidal particles tells us about how they interact with one
another~\cite{fung_measuring_2011}.

These time-series can be big. Although each hologram is a compact
representation of 3D information (a few megabytes), a typical
time-series can consist of tens of thousands of holograms, amounting
to gigabytes of data. The data sets are big because of the way we do
experiments: we use high-speed cameras that can acquire thousands of
frames per second to see 3D motion that is otherwise impossible to
follow.

To handle these large data sets, HoloPy uses the h5py package
(\url{https://www.h5py.org}) to store all holograms and other objects
in compressed HDF5 files~\cite{hdf5}. In addition to reducing memory
requirements, this approach supports lazy loading, so that we can
analyze data sets that are larger than the physical memory of the
computer.

\subsection{Keeping track of metadata}

Another core part of HoloPy is keeping track of metadata. In a typical
experiment, the metadata includes all the information needed to
correctly interpret the hologram, such as what objects were imaged, the
pixel spacing, and the position in a time-series, as well as the
geometry and illumination properties of the microscope. We designed
HoloPy to keep track of this metadata so we didn't have to. HoloPy can
associate metadata with an image when loading it from a file and output
it when saving results. We record the metadata in the same files we use
to record holograms, which must be in a format that supports arbitrary
metadata. We use either HDF5 or TIFF containers.

We also had to figure out how to include metadata in HoloPy's internal
data structures. NumPy arrays cannot handle metadata. We tried to use
Pandas \texttt{DataFrame}, a versatile container for heterogeneous data,
but we found it difficult to adapt to time-series of holograms. We
ultimately decided to use the xarray package~\cite{hoyer2017xarray},
which was designed to handle such multidimensional data sets with
metadata.

Internally, HoloPy stores holograms and metadata in
\texttt{xarray.DataArray} objects. This format has two advantages over a
traditional NumPy array. First, we can assign coordinates to each point
in an image and access parts of the image according to their coordinates
instead of pixel indices. This feature is particularly useful because it
matches how we think about holograms---in terms of regions of space
rather than pixel numbers. Second, we can associate metadata, such as
illumination wavelength and polarization, in the
\texttt{xarray.DataArray.attrs} property, which is a Python dictionary.

During calculations, HoloPy extracts numerical data from the
\texttt{xarray.DataArray} objects and treats it as a simple array, then
repackages the results into an \texttt{xarray.DataArray} before returning
it. With this approach, calculations proceed at the speed of NumPy while
still preserving the metadata. Metadata is also preserved in image
processing tasks such as cropping, background division, or
normalization.

\section{Behind the scenes}

Because experimentalists don't have a lot of time to write code, we made
it easy to adapt HoloPy to new experimental systems or new analysis
techniques. The object hierarchy (Figure~\ref{fig:inference_calc}) makes
it straightforward to add new components such as scattering theories,
scattering models, and inference strategies.

\subsection{Adding new scattering theories}

The core functionality of HoloPy is a universal interface to scattering
models. HoloPy currently contains distinct models for single spheres,
collections of spheres, axisymmetric objects, and arbitrary voxelated
objects. The input to each of these models is the same. Given
descriptions of the object and the camera or detector, HoloPy selects
the scattering model most appropriate for the object and asks the model
to calculate the scattering at the coordinates specified by the
detector. It does this by transforming those coordinates into whatever
coordinate system the model uses. It then adds the scattered field
coherently to the transmitted field to simulate the hologram. Because
the scattered fields from each model are converted to a Python object,
HoloPy can use the same code (a few lines of NumPy array manipulations)
to calculate the hologram from the scattered-field object, irrespective
of the model.

With this modular architecture, we need only define a new class derived
from the \texttt{ScatteringTheory} base class to add a new scattering
model to HoloPy. The new class must define a method that takes inputs of
a HoloPy \texttt{Scatterer} object and an array of coordinates and
returns scattered fields at the coordinate locations. This minimal
requirement gives us the flexibility to wrap scattering codes in
multiple ways. Scattering theories for single and multiple spheres are
compiled into HoloPy modules with the help of f2py. The scattering
theory for axisymmetric objects is compiled into an executable file that
HoloPy runs after writing arguments to a temporary file. And HoloPy
interacts with a command-line interface implementation of the discrete
dipole approximation~\cite{yurkin2011discrete} through Python's
subprocess module. Because the overhead of spawning subprocesses is
small compared to the time required to do scattering calculations, this
approach is convenient and not too costly, except when calculating small
holograms or small subset fractions.

\subsection{Validation}

Since the interface to the scattering codes is a core part of HoloPy, we
must test that it returns correct results. We did not write many of
these codes ourselves, and we must ensure that they compile properly and
return the expected results on different platforms. HoloPy therefore
includes a unit testing framework that tests whether the scattering
models output the correct results for a variety of different conditions.

We've found these tests to be extremely useful when modifying HoloPy,
and particularly when adding a new scattering model. Many such models
can handle spheres as a test case. So, when we added scattering models
for ellipsoids and for clusters of spheres, we were able to validate the
results of these models against our trusted Lorenz-Mie model.

\section{Conclusion}

In our 12 years of working on HoloPy, we've learned a lot about bridging
experiment and computation. For example, we've learned that modularity
encourages new science. Although it's not always easy to wrap a new
scattering code, HoloPy's modularity makes it easy to use once it has
been wrapped. As a result, we can spend more time thinking about what
physical phenomena we want to look at and less about how we'll analyze
the data. For example, after we included a scattering model for
non-spherical colloidal particles, we realized that we could use the
same model for rod-like bacteria. That realization led to an experiment
that revealed some new aspects of bacterial
motion~\cite{wang_tracking_2016}. The experiment wouldn't have happened
had it not been easy to use the same model to analyze two completely
different systems.

We've also found that it's more important for calculations to be
expeditious than efficient. We generally optimize our models only when
it takes more than a few days to analyze the data from an experiment,
because that's the typical timescale of a new experiment. While we're
waiting for the analysis from one experiment, we can do another. If
necessary, we can get results from the analysis more quickly by using
approximate methods like random subsetting. For us, it's more important
that calculations can be set up and started quickly than finished
quickly. A day spent writing code to do the analysis is a day that could
have been spent doing more experiments. But a day that the computer
spends analyzing data doesn't cost us any time in the lab.

That's not to say efficiency isn't important. In fact, figuring out how
to parallelize inference calculations for time-series is one of our
goals for future versions of HoloPy. Although scattering calculations
for a single hologram can be parallelized by pixel using graphics
processing units (GPUs) or multiple processor cores, parallelizing by
frame is more challenging. For a time-series, the best way to get a good
guess for one frame is to use the results of the previous one. As a
result, time-series are more naturally analyzed sequentially than in
parallel. We also have plans to analyze holograms from larger and more
complex objects like living cells. Simulating scattering from these
systems requires complex models with lots of parameters. In both cases,
new methods to find good initial guesses for the parameters could
dramatically speed up the calculations. One potential approach involves
new optimization tools such as the Python package Simple
(\url{https://github.com/chrisstroemel/Simple}). Such techniques can
generate an initial guess quickly by estimating the maximum of the
posterior probability density from a few points in parameter space.

We welcome bug reports, feedback, and contributions to HoloPy. The
source code is licensed under the GNU General Public License version 3
and is hosted on GitHub (\url{https://github.com/manoharan-lab/holopy}).
For convenience, we also distribute binary packages of HoloPy (with all
the scattering codes already compiled) through the \texttt{conda-forge}
channel for the Anaconda Python distribution. The conda package is
cross-platform and is automatically updated whenever a new version is
released. New users can get started with HoloPy by running \texttt{conda
  install -c conda-forge holopy} and following the examples in the
tutorial at \url{http://holopy.readthedocs.io}.

\section*{Acknowledgments}
We acknowledge support from the National Science Foundation through
grant numbers DMR-1306410 and DMR-1420570 and from an Alfred P. Sloan
Research Fellowship.

\bibliographystyle{plain}
\bibliography{holopy}

\end{document}